\documentclass[a4paper,amsmath]{nature}
\usepackage{graphicx}
\usepackage{color}
\newcommand{\dagga}{{\phantom{\dagger}}}
\newcommand{\bk}{\mathbf{k}}
\newcommand{\be}{\begin{equation}}
\newcommand{\ee}{\end{equation}}
\newcommand{\bea}{\begin{eqnarray}}
\newcommand{\eea}{\end{eqnarray}}
\newcommand{\ba}{\begin{eqnarray*}}
\newcommand{\ea}{\end{eqnarray*}}
\title{Kondo Conductance in an Atomic Nanocontact from First Principles}

\author{Procolo Lucignano$^{1,2}$, Riccardo Mazzarello$^{1,3,4}$, Alexander Smogunov$^{1,3,5}$, 
Michele Fabrizio$^{1,3,6}$ \& Erio Tosatti$^{1,3,6}$} 

\begin{document}

\maketitle
\begin{affiliations}
 \item [$^1$] SISSA, Via Beirut 2/4, Trieste 34014, Italy, 
 \item [$^2$] Coherentia CNR-INFM, Monte S.Angelo  via Cintia, 80126 Napoli, Italy, 
 \item [$^3$] INFM, Democritos Unit\'a di Trieste, Via Beirut 2/4, Trieste 34014, Italy,
 \item [$^4$] Computational Science, Department of Chemistry and Applied Biosciences, 
              ETH Zurich, USI Campus, Via Giuseppe Buffi 13, CH-6900 Lugano, Switzerland,
 \item [$^5$] Voronezh State University, University Square 1, Voronezh 394006, Russia, 
 \item [$^6$] ICTP, Strada Costiera 11, Trieste 34014, Italy.
\end{affiliations}




\begin{abstract}
The electrical conductance of atomic metal contacts represents
a powerful tool to detect nanomagnetism. Conductance
reflects magnetism through anomalies at zero bias\cite{knorr2002, zhao2005, neel2007, Brune2008, Vitali2008, 
Ternes2009, reyes2009} -- generally with Fano
lineshapes -- due to the Kondo screening of the magnetic impurity bridging the contact.\cite{hewson,ujsaghy} A full
atomic-level understanding of this nutshell many-body system is of
the greatest importance, especially in view of our increasing need to
control nanocurrents by means of magnetism. Disappointingly, zero bias 
conductance anomalies are not presently calculable from atomistic
scratch. In this Letter we demonstrate a working route connecting approximately
but quantitatively density functional theory (DFT) and numerical renormalization
group (NRG) approaches and leading to a first-principles
conductance calculation for a nanocontact, exemplified by a Ni
impurity in a Au nanowire. A Fano-like conductance lineshape is obtained
microscopically, and shown to be controlled by the impurity $\mathbf{s}$-level
position. We also find a relationship between conductance
anomaly and geometry, and uncover the possibility of
opposite antiferromagnetic and ferromagnetic Kondo screening -- the
latter exhibiting a totally different and unexplored zero bias anomaly. 
The present matching method between DFT and NRG should permit the quantitative 
understanding and exploration of this larger variety of Kondo phenomena at more general 
magnetic nanocontacts.
\end{abstract}

\begin{figure}
\begin{center}
\includegraphics[width=0.8\textwidth]{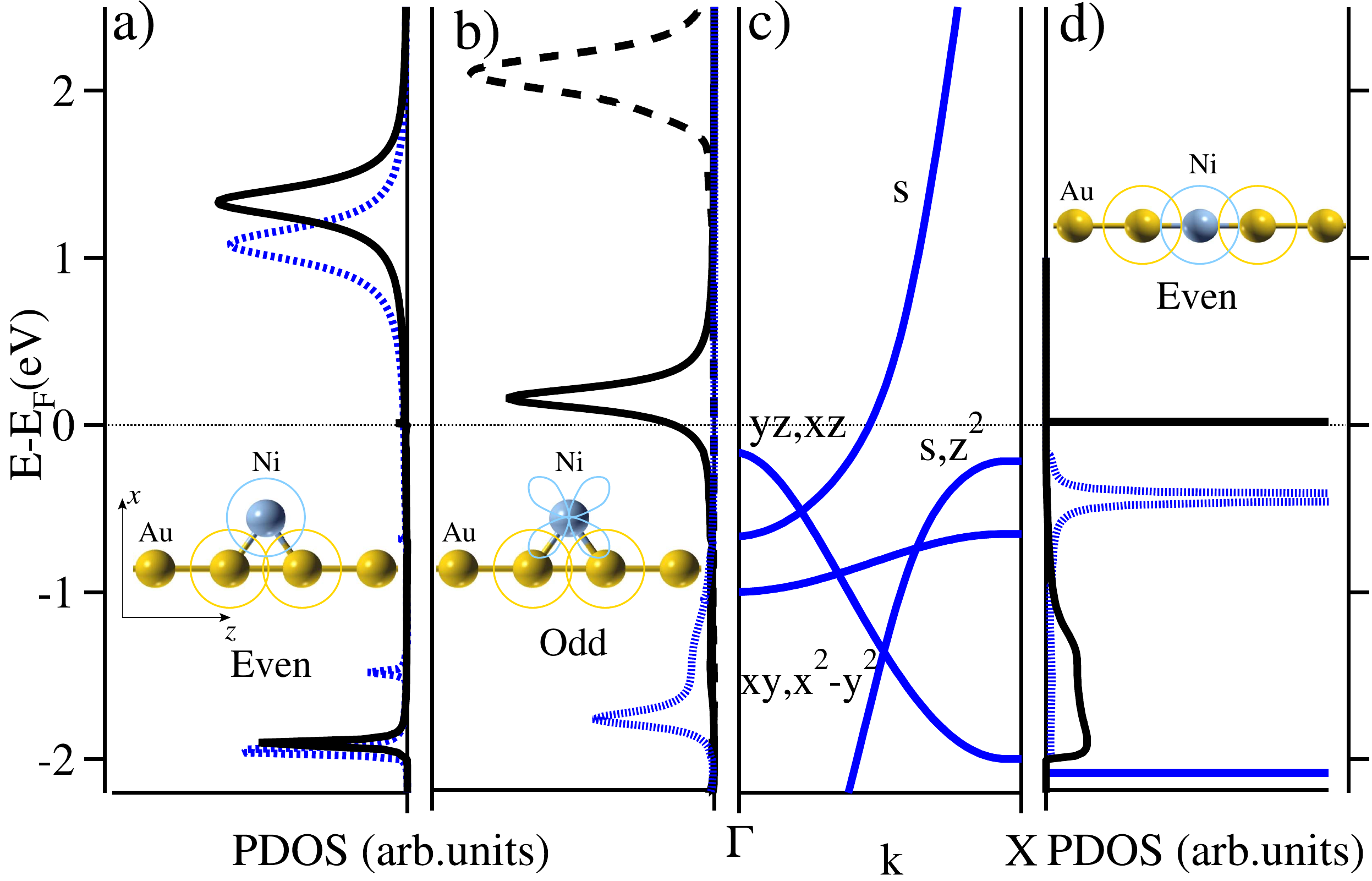}
\end{center}
 \caption{\label{fig1}{Electronic structure of a Au wire with a Ni impurity in bridge and substitutional geometries.} 
a) DFT projected density of states (PDOS) at the Ni $s$ orbital in bridge geometry. 
Blue dotted line: spin up; full black line: spin down. b) PDOS at the Ni $d_{zx}$ orbital in bridge geometry. Black dashed line: 
LDA+U down spin PDOS with $U=3eV$; c) electronic bands of a clean Au chain; d) PDOS at the Ni $d_{zx},d_{zy}$ orbitals 
in substitutional geometry. The effect of $U$ is negligible in this case.}
\end{figure}

The electrical conductance of metals through magnetic nano-contacts, including STM tip
tunnelling\cite{knorr2002, zhao2005, neel2007, Brune2008, Vitali2008,
Ternes2009} and mechanical break junctions (BJ)\cite{reyes2009}
displays, in analogy to quantum dots\cite{kouwenhoven_marcus}, a zero-bias anomaly
reflecting a Kondo many body effect.\cite{hewson,ujsaghy}
First principles calculations of the contact conductance 
and of the anomalies are currently unavailable. The standard  
DFT based Landauer method\cite{1999Choi-PRB} is mean-field in character, 
and thus invalid in presence of many body effects. Alternatively, the NRG\cite{bullaRMP} solution of Anderson impurity 
models (AIM)\cite{Kondo_dot,Ng&Lee} is known to yield correctly the
zero-bias conductance\cite{Meir-Wingreen}: but the AIMs are treated as barely 
more than toy models, their parameters generally guessed or fixed phenomenologically,
rather than based on first principles.

Our approach starts from a first-principles DFT calculation but,
instead of ending as usual with the standard calculation of the mean-field 
conductance\cite{bagrets2007,miuraPRB2008}, it continues with the building of 
geometry-dependent AIMs, each explicitly including the channels, orbitals, and especially symmetries  
suggested by the DFT electronic structure. The key point is that the numerical 
parameters are fine tuned by requiring these AIMs to yield in mean field 
the same phase shifts suffered in DFT by a spin polarized traveling electron 
of each symmetry. We exemplify directly our method by straightforward 
application to a prototype Au-Ni-Au monatomic chain contact, a test system 
chosen because of its simplicity: in fact, Au possesses just one $s$-like 
conduction band, and the Ni bridging impurity is free of complications involving 
anisotropy and spin-orbit coupling, which affect other transition metal atoms.\cite{Brune2008} 

We start off from the recent DFT calculation of the nanocontact electronic 
structure\cite{miuraPRB2008} and examine the symmetry-selected projected density of 
electron states (PDOS). The Ni atom in the Au chain (axis parallel to $z$) 
has two low energy geometries: bridge (B) and substitutional (SUB) 
(left and right insets in Fig.~\ref{fig1}). 
In B (preferred at zero stress) the PDOS indicates a well defined Ni 3$d_{zx}$ spin down empty 
orbital (Fig.\ref{fig1}b). The Au chain conduction channel is 
a single $6s$ nondegenerate band crossing the Fermi energy $E_F$ (Fig. \ref{fig1}c), hybridized 
with both Ni $4s$ and Ni $3d_{zx}$ orbitals -- the first even under reflection 
across the Ni $xy$ plane, the other odd. Electrons traveling in the Au ``leads''
will thus scatter differently off the Ni impurity 
in the two symmetry channels, even ($e$) and odd ($o$). 
The DFT calculated magnetic moment $M=2\mu_B \langle S_z \rangle$ is  
$1.30\mu_B$, enhanced over the bare value of a $S=1/2$ impurity state 
by co-polarization of nearby Au atoms, but dropping in fact to $\sim 1.0\mu_B$ after
correction of DFT through a Hubbard-$U$ term 
(so-called LDA+U) which raises the $3d_{zx}$ energy\cite{miuraPRB2008} 
(dashed curve in Fig.\ref{fig1}b). 
Summing up, the B geometry implies electron hopping across Ni $3d_{zx}$ which has odd symmetry  
and $S\sim 1/2$, as well as across Ni $4s$ which has even symmetry and is very little spin
polarized.
In SUB geometry (of higher total energy\cite{miuraPRB2008}, but probably accessible at large stress), 
the electron states can be labeled by $m$, the orbital angular momentum 
about the $z$ axis. The empty spin-down Ni state orbitals are now 
a $3(d_{zx},d_{zy})$ degenerate pair with $|m|=1$.
Orthogonality to the $m=0$ Au conduction band 
results in a PDOS $\delta$-like peak just 0.02~eV above $E_F$ (Fig.~\ref{fig1}d). 
This suggests $S=1$; and the DFT calculated Ni magnetic moment is in fact
$M=1.91 \mu_B$, rising to $2.02\mu_B$ upon addition of a $U$ term.\cite{miuraPRB2008}
Summing up, in the SUB geometry the Ni impurity has $S\simeq 1$ and the magnetic orbitals are ``spectators''. 
The Au conduction electrons hop onto the Ni $4s$ orbital in the even symmetry channel, whereas they find no
Ni valence orbital in the odd channel.

The DFT calculations also provide the electron transmission matrix
through the impurity, yielding directly the Landauer ballistic conductance.
However, in standard DFT the moment is artificially frozen, breaking spin rotational 
invariance, thus wrongly predicting a different conductance for spin up and spin 
down electrons\cite{miuraPRB2008} -- 
in zero external field there simply is no telling what is up and what is down. 
Through the Kondo effect,\cite{hewson} 
symmetry is restored by screening of the spin by the conduction 
electrons. This many body effect is usually studied by applying
a  viable spin-symmetric technique such as the NRG\cite{bullaRMP}  
to a microscopic, but phenomenological AIM. This wastes the detailed
information provided by DFT, unless some way is found to preserve it.
The extraction of Anderson model parameters from DFT has a long history.\cite{gunnarsson89}  
Very recent work on Fe impurities in bulk Au and Ag also addressed this problem.\cite{Costi2009}
Our proposed joining step between DFT and NRG is the construction of a different, 
specifically designed Anderson 
model for each symmetry and geometry, with parameters embodying quantitatively 
the DFT calculated electron-impurity scattering amplitudes and phase shifts. 
For each symmetry ($e$ or $o$) 
an incoming electron wave with spin initially along $x$ will, after scattering 
on Ni (magnetic moment parallel to the $z$-axis), rotate in the $x-y$ plane, 
clockwise if the coupling is FM or counter-clockwise if it is AFM.
Within DFT, electron-impurity scattering 
leads to calculable spin rotation angles, $\theta_{e/o} = 2(\delta_{e/o}^\downarrow-\delta^\uparrow _{e/o})$,   
which measure the phase shifts and thus the coupling sign and strength of 
the Ni impurity spin to the Au conduction electrons. 
The DFT calculated rotation angles (Table~\ref{table1}) show that 
in geometry B the coupling is large and AFM in the odd channel, small and FM in the 
even channel. In geometry SUB the coupling is on the contrary weakly FM in both channels.

\begin{table}[t]
\begin{center}
\begin{tabular}{ccc} \hline\hline
 Angle (rad)             &   Bridge DFT (HF-AIM) & Subst. DFT (HF-AIM) \\ \hline
$\theta_{e}$             &  -0.28  (-0.28)  &  -0.19 (-0.19) \\
$\theta_{o}$             &  1.10  (1.17)    &  -0.12 (-0.00) \\ \hline\hline
\end{tabular}
\end{center}
\caption
{{\bf Deflection angles.} Spin rotation angles $\theta_{e/o}$ (at the Fermi energy) of an 
incoming transverse-polarized conduction electron wave after scattering at the Ni impurity 
for even and odd conduction channels. 
DFT ab initio angles are compared with the AIM Hartree Fock ones (in brackets),
optimized by adjusting the AIM parameters.
Positive $\theta_{e/o}$ correspond to 
antiferromagnetic electron-impurity coupling, negative to ferromagnetic. 
$\theta_{o}$ substitutional is zero in the AIM model, 
as there is no Ni empty valence orbital of this symmetry, and inter-site 
exchange is neglected.}
\label{table1}
\end{table}

%
\begin{figure}
\begin{center}
\includegraphics[width=0.8\textwidth]{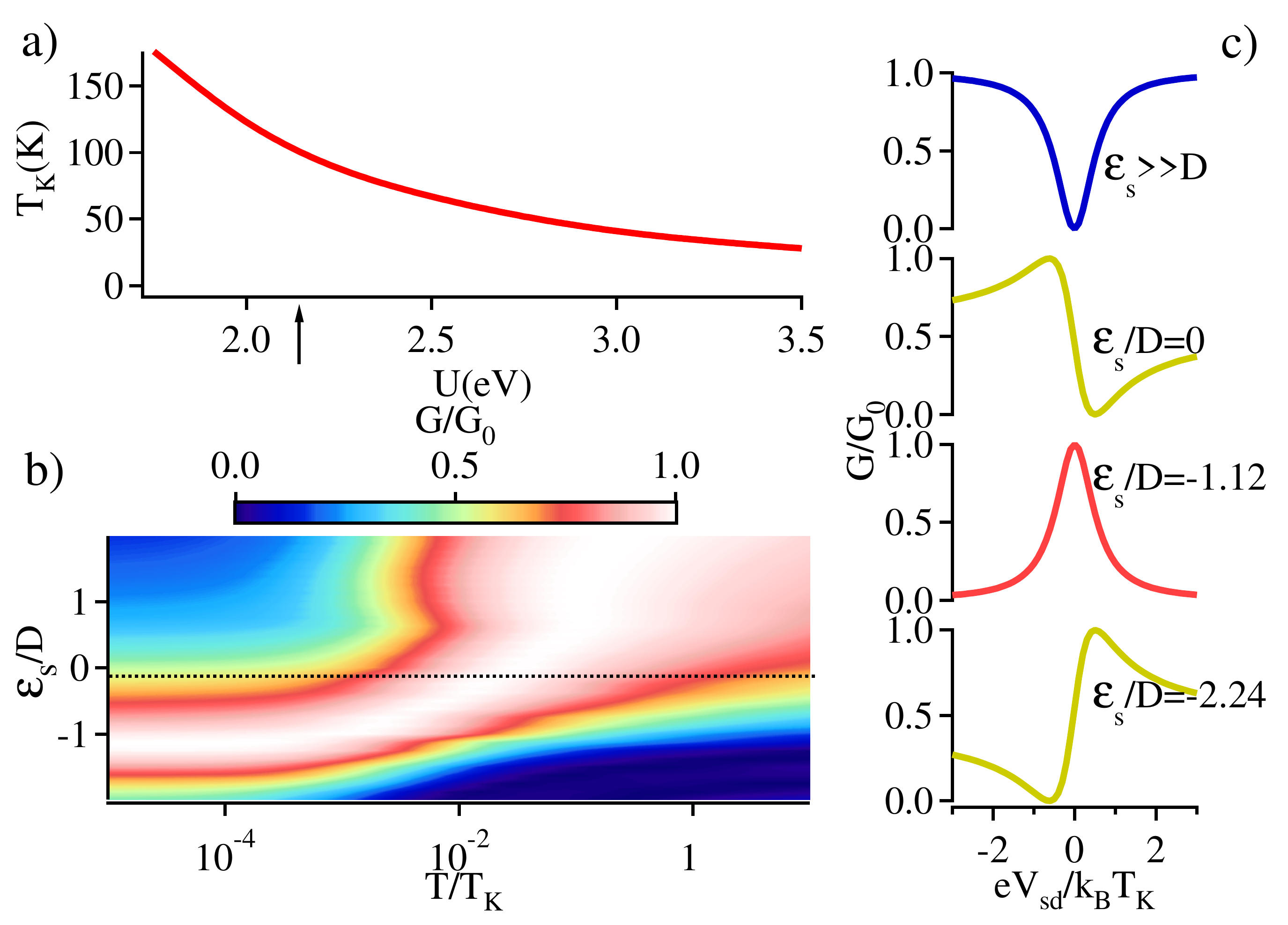}
\end{center}
 \caption{\label{fig2}{Conductance properties in the bridge geometry}. a) NRG calculated Kondo temperature $T_K$ 
illustratively shown as a function of the on-site $d_{zx}$ interaction $U$.
The physical value for Ni 3$d_{zx}$ is indicated by the arrow. 
b) Color plot of the zero bias conductance as a function of temperature 
(normalized to the Kondo temperature) and of the effective $s$ 
orbital energy $\epsilon_s/D$ treated as a parameter for illustrative 
purposes. The physical value for Ni 4$s$ is indicated by the horizontal line. 
When $\epsilon_s >>E_F$ the low temperature  ($T<T_K$) conductance 
is zero, and increases by moving $\epsilon_s$ towards $E_F$, hitting a maximum  
at $\epsilon_s\sim-D$, declining below -$D$. c) Low temperature differential conductance as a function of bias voltage $V_{sd}$  
applied to the junction.  The color of curves refer to the value of 
$\epsilon_s$ in the left bottom panel. Note the Fano lineshapes, due 
to variation of the even-odd channel interference 
with varying $\epsilon_s$. (See supplementary material).}
\end{figure}

We now have all the ingredients needed to construct the minimal AIMs 
which describe the Au-Ni-Au contact. The models must include the Ni $4s$ 
orbital, with energy $\epsilon_s$ and creation operator $s^\dagger_\sigma$, and 
the $3d_\alpha$ orbitals, with energies $\epsilon_\alpha$ and creation operators 
$d^\dagger_{\alpha\sigma}$, where $\sigma$ denotes spin and $\alpha=xy,xz,\dots$ 
labels orbital symmetry. Electron transport in the Au leads takes place in the $6s$ 
band (the $5d$ band is filled)
leading to Au conduction state combinations with $p=e,o$  spin-$\sigma$ creation operators 
$c^\dagger_{\bk p \sigma}$ and energy $\epsilon_{p\bk}$. 
The AIM Hamiltonians have the general form   
\bea
H &=& \sum_{\bk p\sigma}\,\epsilon_{p\bk}\,c^\dagger_{\bk p\sigma} c^\dagga_{\bk p\sigma}
+\sum_{\bk \sigma} V_{\bk s}\, \left(c^\dagger_{\bk e \sigma} s^\dagga_{\sigma} + H.c.\right)\nonumber\\
&&+ \sum_{\bk p \sigma}\, \sum_{\alpha}\, V_{\bk p \alpha}\, \left(c^\dagger_{\bk p \sigma} d^\dagga_{\alpha \sigma} + H.c.\right)
\label{Ham}\\
&&+ \sum_\sigma \epsilon_s\,s^\dagger_\sigma s^\dagga_\sigma 
+ \sum_{\sigma\alpha}\,\epsilon_\alpha\,d^\dagger_{\alpha\sigma}d^\dagga_{\alpha\sigma} + H_{int}\left[s,d_\alpha\right].\nonumber
\eea
In (\ref{Ham}), $V_{\bk s}$ and $V_{\bk p \alpha}$ are hybridization parameters
between leads and impurity orbitals,
while $H_{int}\left[s,d_\alpha\right]$ includes the mutual interactions between 
impurity orbitals. All parameters in (\ref{Ham}) depend on the geometry of 
the contact, and must reflect the physics provided by the DFT calculation. 
Our practical procedure is to estimate first approximate AIM parameters from direct
inspection of the DFT electronic structure PDOS,
and to operate successively a finer adjustment by comparison of the Hartree-Fock
calculated AIM spin rotation angles with those directly provided by DFT. Because
the AIMs include only a few orbitals and interactions, the agreement is 
semi-quantitative: yet it can, as we will show, be made very significant.  

{\sl Bridge} -- 
In B geometry, the Ni 3$d_{zx}$ orbital is directly hybridized with the odd 
conduction states, the Ni 4$s$ with the even states, and all other Ni 3$d$ states
can be dropped. The interaction $H_{int}$ 
includes a Hubbard $U\sim 2\mathrm{eV}$ (deduced by the exchange splitting 
of spin minority and majority 3$d_{zx}$ PDOS peaks ), and an intra-atomic $s$-$d$ FM exchange   
$J_H\sim -0.3eV$ (extracted by the spin splitting of Ni 4$s$, 
co-polarized by the magnetic 3$d_{xz}$). A band-impurity hybridization width 
$\Gamma_{zx} = 2\pi \rho_0 V^2_{xz} \simeq 0.25~eV$  
($\rho_0\sim 0.1 eV^{-1}$ is the conduction band density of states at $E_F$)
is deduced from the broadening of the DFT calculated 
PDOS Ni 3$d_{zx}$ peaks (see Fig.\ref{fig1}b). Moreover, the majority and minority peak 
asymmetry around $E_F$ is strongly reduced by addition of a realistic Hubbard $U$ 
in LDA+U (see dashed peak in Fig.\ref{fig1}b), so 
we may assume a particle-hole symmetric $\epsilon_{zx}=-U/2$, with $U \sim2 - 3$~eV. Finally, Ni 4$s$ is a broader orbital, 
for which $U\sim 0$ and $\Gamma_s \sim 3$~eV, as extracted 
from the PDOS linewidth. The $s$-level energy parameter $\epsilon_{s}$ 
is at the outset totally uncertain. The fine adjustment to reproduce the 
DFT spin rotation angle yields $\epsilon_{s} = -0.4~eV $, just below $E_F$. 
For the sake of illustration, since $\epsilon_s$ will sensitively control the Fano 
lineshape of the conductance anomaly, we choose nevertheless to explore different values of $\epsilon_{s}$, 
both above and below $E_F$, as might be relevant in other nanocontacts. Note 
that the bridge configuration has an additional direct Au-Au 
hopping bypassing the impurity (see inset of Fig\ref{fig1}b), its effect
taken equal to $\Gamma_s$ (for further detail see supplementary).

The AIM Hamiltonian (\ref{Ham}) with the parameters adjusted as above 
yields in the HF approximation spin rotation angles quite close to those 
calculated by fully realistic DFT (Table~\ref{table1}, see also supplementary material),
so that the low energy description of the B configuration 
is close to being quantitatively correct. The next step is the NRG treatment of model (\ref{Ham}) 
(see Methods). At low temperatures, $J_H$ becomes irrelevant and the $e$ and $o$ channels behave 
independently. The conductance $G=G_0\sin(\delta_e-\delta_o)^2$, where $G_0=2e^2/h$, is 
calculated by following, during the NRG flow, the phase shifts $\delta_{e/o}$ (see Methods).
The competition between the Kondo effect in the strongly AFM odd channel and the resonance 
in the weakly FM even channel results in the zero bias conductance plotted in 
Fig.~\ref{fig2}b as a function of temperature and of $\epsilon_s$, the conventional 
position of Ni 4$s$ relative to $E_F$. At large unphysical $\epsilon_s >>D$ 
($D=2.5eV$ is half bandwidth),  
both channels acquire a $\pi/2$ phase shift, with a destructive interference and  
zero conductance. At the fitted value of $-D\ll \epsilon_s < 0$, the conductance is  
instead expected to be $G/G_0 \sim 0.5$. This deviates considerably from the DFT 
calculated Landauer conductance, relatively close to the unitary limit 
in both spin channels, $G = G_{up} + G_{down} \sim 0.5  + 0.4 \sim 0.9G_0$.\cite{miuraPRB2008} 
The large difference underscores, as anticipated, the impact 
of Kondo correlations, ignored by DFT.
By further decreasing $\epsilon_s<0$, the conductance reaches its maximum value $G/G_0=1$ 
and then drops. This effect of $\epsilon_s$ can be assimilated to that of a gate voltage, 
controlling the conductance lineshape versus source-drain voltage. 
Following Ref.~\citen{Meir-Wingreen} we  
calculated the nonequilibrium conductance as a function of the source-drain 
voltage $V_{sd}$ (details given in supplementary material).
The conductance shows a  Fano-like resonance near zero bias,
as shown in Fig.~\ref{fig2}c, and as expected on general grounds.\cite{ujsaghy}     
The large predicted Kondo temperature $\sim 100~K$, see Fig.~\ref{fig2}a,  
should make this anomaly readily accessible in a break-junction transport experiment. 

\begin{figure}
\begin{center}
\includegraphics[width=0.8\textwidth]{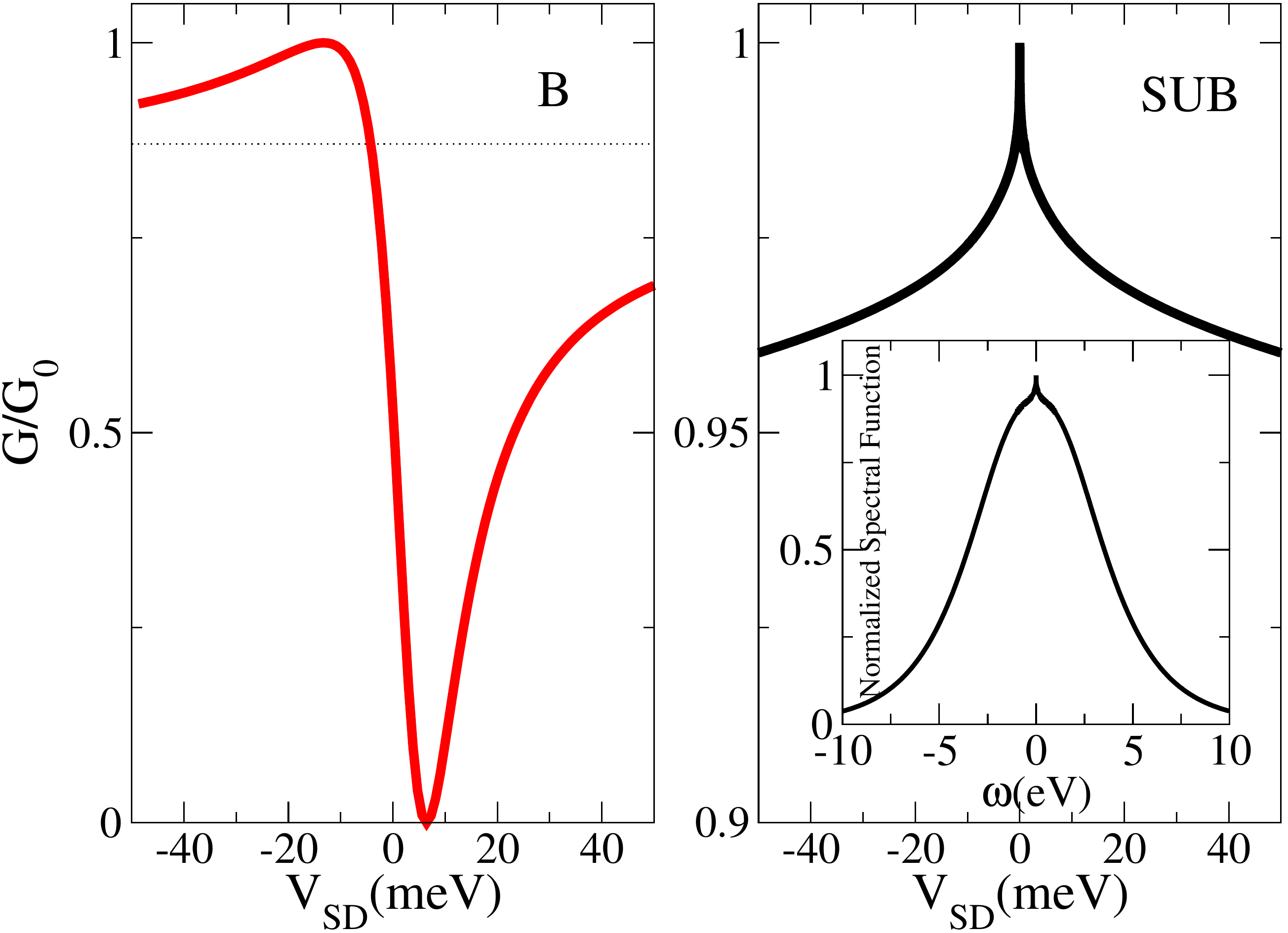}
\end{center}
\caption{\label{sketch}{Conductance vs. source-drain voltage of regular and of Ferro Kondo in the two geometries.} 
Sketch of the different zero bias anomalies expected for the current through
an Au-Ni-Au junction in the B (left panel), with realistic $\epsilon_s=-0.4~\mathrm{eV}$, and SUB (right panel) geometry.
Opposite to the Fano lineshape calculated in the B geometry (see also Fig.\ref{fig2}c), 
the Kondo antiscreening in SUB geometry yields a spectral function (see inset) which, for small energies, 
exhibits a logarithmic ``pimple'' (see text).}
\end{figure}

{\sl Substitutional} -- In SUB geometry, magnetism occurs  
in 3$d_{zx}$ and 3$d_{zy}$ Ni orbitals that do not hybridize with 
the Au 6$s$ conduction band.
The interaction term $H_{int}\left[s,d_\alpha\right]$ in Eq.~(\ref{Ham}) 
should now include all multiplet exchange splittings between 3$d_{zx}$ and 3$d_{zy}$ 
degenerate magnetic orbitals. The intra-atomic ferromagnetic exchange 
with the 4$s$ Ni orbital is weak, see the small DFT rotation angles in Tab.\ref{table1},
and all the electron-impurity phase shifts are now FM. 
Unlike regular AFM Kondo, the ferromagnetic Kondo exchange 
is well known to be irrelevant at low temperatures,~\cite{hewson} where the magnetic orbitals 
asymptotically decouple from the conduction band.
Only Au 6$s$ and Ni 4$s$ hybridization will effectively survive, leading to a $\pi/2$
phase shift in the even channel. The  zero bias 
conductance will thus approach unitarity $G(V=0) \le G_0$, with
a broad Breit-Wigner lineshape of width $\Gamma_s$ expected for the spectral function.
While confirming that, the explicit NRG calculation with the Anderson model 
(1) with orbitals and parameters appropriate to the SUB geometry
shows in addition that the FM Kondo antiscreening mechanism gives 
rise at the Fermi level to a logarithmically sharp spectral function``pimple'', see Fig.~\ref{sketch} 
(right panel) and also Ref.~\citen{Koller2005}, pushing 
$G/G_0$ up to unity at zero bias. This is again different from the DFT
Landauer mean field conductance,\cite{miuraPRB2008} 
$G_\uparrow + G_\downarrow \sim (0.40 + 0.38)G_0 \sim  0.8 G_0$. 
The lineshape difference of this expected ``ferro Kondo'' anomaly in SUB 
geometry and of the Fano lineshape of ordinary Kondo in B geometry, respectively right and left panels in Fig.~\ref{sketch}, 
is striking. More generally, no ferro Kondo zero bias anomaly appears to have been reported so far anywhere,
so its existence in the present or in other nanocontacts is open for
experimental detection. 
While more work will be needed to describe the effect of a magnetic field
in a ferro Kondo case, the effect of temperature should be to cut off
the logarithmic cusp, modifying the conductance in the form $1 - const/\ln^2 T$.
The final conceptual aspect of spin rotational symmetry is worth mentioning.
As is well known,\cite{nozieres} in the regular AFM Kondo, for example of our B geometry, 
the impurity approximate $S=1/2$ is absorbed by Kondo screening into an exact singlet, 
a trivially rotationally symmetric state. In the FM Kondo coupling of SUB geometry
the impurity's approximate $S=1$ is turned by Kondo {\it antiscreening} into an exact 
triplet, fully SU(2) rotationally symmetric, endowed with 
its $2S+1$ degeneracy. 

Summarizing, this work represents a first attempt at joining together the quantitative 
advantages of DFT electronic structure with the correct NRG many body Kondo physics of a magnetic impurity
in a break junction metal contact. Because the intermediate Anderson models can only 
represent approximately the full electronic processes, the procedure is
semi-quantitative. It is nonetheless revealing, showing in the regular Kondo 
case an \emph{ab initio} example of Fano lineshape; involving phase shifts 
and Kondo processes that are sometimes ferromagnetic and involve antiscreening;
and predicting very different zero bias anomalies depending on impurity geometry and spin. 
The method should be applicable more generally, whenever 
well defined intermediate AIMs a) can be built and b) are soluble.  
On the experimental side, 
numerous magnetic atom nanocontacts have been realized
using STM\cite{neel2007, Brune2008, Vitali2008,
Ternes2009}. 
Mechanical break-junction studies have just begun, so far exploring a different 
type of phenomenon, namely the emergence of Kondo effects in ferromagnetic 
metals\cite{reyes2009}, but many more should be possible in the future, when the 
kind of antiferro-ferro Kondo switching described here should hopefully be pursued.

\pagebreak
\section*{METHODS}
\subsection{DFT}
We carried out DFT calculations within the 
spin-polarized generalized-gradient approximation ($\sigma$-GGA) with the 
Perdew-Burke-Ernzerhof parametrization~\cite{1996Perdew-PRB} for the 
exchange and correlation energy, with same parameters as in Ref.\citen{miuraPRB2008}.
All the calculations were done with the plane wave {\tt PWscf} code, 
included in the {\tt QUANTUM-Espresso} package.~\cite{espresso} 
Ultrasoft  pseudopotentials~\cite{1990Vanderbilt-PRB} were employed and
kinetic energy cutoffs of $30$ Ry and $300$ Ry were used for the wave functions 
and the charge density, respectively.
For both B and SUB geometries, the Au-Au bond length was taken $2.80$ \AA 
to avoid spurious magnetizations due to DFT self-interaction errors,~\cite{miuraPRB2008} 
whereas all the Au-Ni distances were fully optimized. 
The calculation supercell consisted of $16$ Au atoms plus one Ni atom periodically 
repeated in all three directions. The wire-wire distance in the $xy$ plane (perpendicular to the wire axis $z$) 
was $10.58$ \AA, making spurious interactions between periodic replicas negligible. 
In the B configuration, the Ni atom lies in the $xz$ plane. 
Convergence with respect to $k$-points and smearing parameters was carefully checked.

Transmission and reflection amplitudes, and the spin rotation angles (see the Supplementary material) 
were calculated using the Choi and Ihm's method~\cite{1999Choi-PRB} generalized 
to ultrasoft pseudopotentials,~\cite{2004Smogunov-PRB} as implemented in the PWCOND code 
(a part of the {\tt QUANTUM-Espresso} package). The self-consistent potential in the first part of the supercell 
described above, of length equal to the Au-Au distance, was used to build the periodic potential 
of the left and right leads, while the potential in the rest of the supercell was used as the scattering region. 
The zero bias ballistic conductance was obtained using the Landauer-B\"uttiker formula~\cite{Landauer}
from the transmission coefficient at $E_F$ (with all spin moments 
frozen). Spin-orbit effects were not taken into account
in the present study. The PDOS in Fig. \ref{fig1} were calculated directly from the scattering states.

\subsection{NRG}

In our NRG code (for a review see Ref.~\citen{bullaRMP}) we implement the $U(1)$ charge symmetry 
(quantum number $Q$, the total charge with respect to one electron per site and orbital) and the $SU(2)$ spin symmetry 
(quantum number $S$, the total spin). 
We choose the Wilson discretization parameter $\Lambda=2$ (or $1.8$ when calculating the Green functions), and  
keep up to 1500 states per iteration when calculating dynamical quantities, or 800 if we are only interested in the energy spectrum.
The Kondo temperature can be expressed as\cite{hewson}  $T_K= \frac{\pi w Z \Gamma}{4  k_B}$ where $w=0.4128$ 
is the Wilson coefficient, $\Gamma$ the hybridization linewidth in the magnetic channel, and $Z$ the quasiparticle residue. 
The latter can be extracted from the self-energy according to:
$
Z^{-1}=\left.1-({\partial \Sigma(i\omega_n)}/{\partial \,i \omega_n})\right|_{i\omega_n=0}\:.
$
The self-energy is $\Sigma=-G^{-1}+G_0^{-1}$ where $G$ and $G_0$ are the impurity Green functions calculated by NRG, respectively 
in the presence and absence of interaction.  The resulting Kondo temperature in our model is shown in Fig.~\ref{fig2} 
as a function of  $U$.
In the bridge geometry the zero bias conductance is evaluated  as a function of temperature by direct inspection of the NRG flow.
It can be expressed as $G=2e^2 \,\sin^2(\delta_{e}-\delta_{o})/h$. The phase shift $\delta_{e/o}$ are related to 
the two lowest energies of states with quantum numbers $(Q,S)=(1,1/2)$, which correspond to the cost of adding an even or odd 
electron to the ground state that has quantum numbers $(0,0)$.
We calculated the difference between these phase shifts, hence the zero-bias conductance,  as a function of the temperature $T$ extracted from the NRG iterations. 
We note however that, while we are quite confident about the values at low temperatures,  those at high temperatures must be taken 
with caution, since the spectrum is still far from a Fermi liquid one.
In the substitutional geometry we calculate the Ni $4s$ spectral function following the guidelines of 
Ref.\citen{costi_hewson_zlaticJPC94}.




\begin{addendum}
 \item[Acknowledgments] We would like to thank D. Basko and C. Untiedt for very useful
discussions. The work was supported by the Italian Ministry of University
and Research, through a PRIN-COFIN award, and by INFM through ``Iniziativa 
Trasversale Calcolo Parallelo''. The environment provided by the
independent ESF project CNR-FANAS-AFRI was also useful. P.L. acknowledges financial support from EC STREP project MIDAS  
``Macroscopic Interference Devices for Atomic and  Solid State Physics''   
and CNR-INFM within ESF Eurocores Programme FoNE-Spintra. 
\item[Competing Interests] The authors declare that they have no competing financial interests.
\item[Correspondence] Correspondence to: Erio Tosatti$^{1,3,6}$. Correspondence and requests for materials
should be addressed to E.T.~(Email: tosatti@sissa.it).
\item[Author Contributions] P.L., M.F. and E.T. conceived and elaborated the Kondo aspects, including NRG; R.M. and A.S. 
worked out the DFT part, from which A.S. extracted the phase shifts. E.T. wrote the paper, with help from all co-authors. 
 
\end{addendum}

\end{document}